\documentclass[aps,prl,preprint,superscriptaddress,showpacs]{revtex4}

\usepackage{graphicx}
\usepackage{bm}        
\usepackage{amsmath}
\usepackage{amssymb}   

\def\k{{\bm k}}

\newcommand{\beq}{\begin{equation}}
\newcommand{\eeq}{\end{equation}}

\newcommand{\crea}[3]{\hat{#1}^{\dagger}_{#2,#3}}
\newcommand{\anni}[3]{\hat{#1}_{#2,#3}}

\begin{document}

\title{External field control of collective spin excitations in an optical lattice of $^2\Sigma$ molecules}

\author{J. P\'{e}rez-R\'{\i}os}
\affiliation{Department of Chemistry, University of British Columbia, Vancouver, V6T 1Z1, Canada}
\affiliation{Instituto de F\'{i}sica Fundamental, Consejo Superior de Investigaciones Cient\'{i}ficas,Ê Serrano 123, 28006 Madrid, Spain}
\author{F. Herrera}
\affiliation{Department of Chemistry, University of British Columbia, Vancouver, V6T 1Z1, Canada}
\author{R. V. Krems}
\affiliation{Department of Chemistry, University of British Columbia, Vancouver, V6T 1Z1, Canada}

\date{\today}

\begin{abstract}

We show that an ensemble of  $^2\Sigma$ molecules in the rotationally ground state trapped on an optical lattice exhibits collective spin excitations that can be controlled by applying superimposed electric and magnetic fields. In particular, we show that the lowest energy excitation of the molecular ensemble at certain combinations of electric and magnetic fields leads to the formation of a magnetic Frenkel exciton. The exciton bandwidth can be tuned by varying the electric or magnetic fields.  We show that the exciton states can be localized by creating vacancies in the optical lattice. The localization patterns of the magnetic exciton states are sensitive to the number and distribution of vacancies, which can be exploited for engineering many-body entangled spin states. 
We consider the dynamics of magnetic exciton wavepackets and show that the spin excitation transfer between molecules in an optical lattice can be accelerated or slowed down by tuning an external magnetic or electric field.

\end{abstract}


\maketitle

\section{Introduction}

A major thrust of current experimental research is to create ultracold ensembles of polar molecules trapped on an optical lattice \cite{carr,book}. Optical lattices of ultracold molecules  are predicted to be ideally suited for quantum simulation of complex many-body quantum systems \cite{micheli, wang2006, buchler2007} and the development of new schemes for quantum information storage and processing \cite{demille, rabl2006}. These applications exploit (i) the possibility to produce molecular ensembles in a Mott-insulator state, i.e. an ordered array with one molecule per lattice site \cite{Greiner2002,Danzl2010}; (ii) the presence of long-range dipole - dipole interactions that can be used to couple molecules in different lattice sites \cite{Goral2002}; (iii) the rotational structure of molecules; (iv) the possibility to address spectroscopically molecules in specific sites of an optical lattice  \cite{demille}. For example, DeMille proposed to use rotational states of polar molecules confined in a one-dimensional optical lattice as qubits of a quantum computer entangled by the dipole - dipole interactions \cite{demille}. Recently, Micheli and coworkers showed that ultracold molecules in the $^2\Sigma$ electronic state trapped in a two-dimensional optical lattice can be used for engineering lattice-spin models that give rise to topologically ordered states \cite{micheli}.  This stimulated experimental work on the creation of ultracold molecules in electronic states with unpaired electrons \cite{carr}. 

  Ultracold $^2\Sigma$ molecules can be produced by photoassociation of ultracold alkali metal atoms with ultracold alkaline earth  \cite{Jones2006}  or closed shell lanthanide atoms \cite{Okano2009,Nemitz2009}, buffer gas loading \cite{doyle-cah,doyle-caf} or direct laser cooling \cite{dirosa}, as was recently demonstrated for the molecule SrF \cite{Shuman2009}.  The presence of the unpaired electron in a $^2\Sigma$ molecule allows for new applications of ultracold molecules exploiting weak couplings of the electron spin with the rotational angular momentum of the molecule \cite{carr,book,micheli}. In particular, the electron spin of $^2\Sigma$ molecules can be used for encoding quantum information as in other spin-1/2 particles \cite{book-on-quantum-information} and for quantum simulation of many-body spin dynamics \cite{micheli}. In order to realize these applications, it is necessary to develop techniques for entangling the spin degrees of freedom and controlled preparation of many-body spin-dependent states of ultracold molecules on an optical lattice.  Micheli and coworkers showed that the electric dipole - dipole interaction between $^2\Sigma$ molecules on an optical lattice leads to spin-dependent binary interactions whose parameters can be tuned by a combination of dc electric and microwave fields \cite{micheli}. 
Here, we extend the work of Micheli and coworkers to explore the possibility of tuning the dynamics of collective spin excitations in an ensemble of $^2\Sigma$ molecules on an optical lattice by external electric and magnetic fields. 

  We consider SrF($^2\Sigma$) molecules in the ro-vibrationally ground state confined in a Mott-insulator state in an optical lattice. In the presence of a weak magnetic field, the lowest energy excitation of the molecular crystal corresponds to the spin-down to spin-up transition in an isolated molecule. We show that for certain combinations of superimposed electric and magnetic fields, the lowest energy excitation of the molecular crystal leads to the formation of a magnetic Frenkel exciton. The magnetic exciton is a many-body entangled state of the molecules. If some of the molecules are removed from the optical lattice to produce vacancies, which can be achieved by a recently demonstrated technique \cite{vacancies}, the exciton undergoes coherent localization due to scattering by the impurities. Using a series of examples for SrF molecules in an optical lattice, we show that the localization of the magnetic exciton can be controlled by varying the concentration and distribution of vacancies as well as the external magnetic or electric fields. The system proposed here can be used for quantum simulation of spin excitation transfer in many-body crystals without phonons. We show that the spin excitation transfer between molecules can be accelerated or slowed down by tuning an external magnetic field.

\section{Theory}

\subsection{Energy levels of a $^2\Sigma$ molecule}

The Hamiltonian of an isolated $^2\Sigma$ molecule in the presence of superimposed electric and magnetic fields can be written as  \cite{mizushima,roman-jcp2004,timur-prl2006}

\beq\label{e2}
\hat{H}_{\text{as}} = \hat{H}_{\rm ro-vib} + \gamma_{\rm SR} {\bm S}\cdot{\bm N} - {\bm E}\cdot{\bm d} + 2\mu_{\text{B}}{\bm B}\cdot{\bm S},
\eeq
where the first term determines the ro-vibrational structure of the molecule, $\gamma_{\rm SR}$ is the constant  of the spin-rotation interaction between the rotational angular momentum ${\bm N}$ and the spin angular momentum ${\bm S}$ of the molecule, ${\bm E}$ and ${\bm B}$ are the vectors of the electric and magnetic fields, ${\bm d}$ is the dipole moment of the molecule and $\mu_B$ is the Bohr magneton. We assume that both ${\bm E}$ and ${\bm B}$ are directed along the quantization axis $z$. We consider molecules in the vibrationally ground state and use the rigid-rotor approximation for our calculations. 
It is convenient to use the basis of direct products \cite{roman-jcp2004} of the rotational $|N M_N \rangle$ and spin $| S M_S \rangle$ wave functions to evaluate the eigenvectors and eigenvalues of Hamiltonian (\ref{e2}). Here, $M_N$ and $M_S$ denote the projections of ${\bm N}$ and ${\bm S}$, respectively, on the $z$ axis.  
The diagonalization of Hamiltonian (\ref{e2}) yields the energy levels of the molecule in superimposed electric and magnetic fields, shown in Figure 1 for the particular example of SrF. The eigenvectors of Hamiltonian (\ref{e2}) are linear combinations

\beq\label{eigenvectors}
\phi_f = \sum_{N M_N} \sum_{M_S} C_{N M_N M_S}^f |N M_N \rangle |S M_S \rangle
\eeq

\noindent
The index $f$ denotes the $f$-th excited state of the molecule. We neglect the hyperfine structure of the molecule. This is a good approximation for the magnetic fields considered in the present work.

\begin{figure}[ht]
\label{energy-levels}
\begin{center}
\includegraphics[scale=0.55]{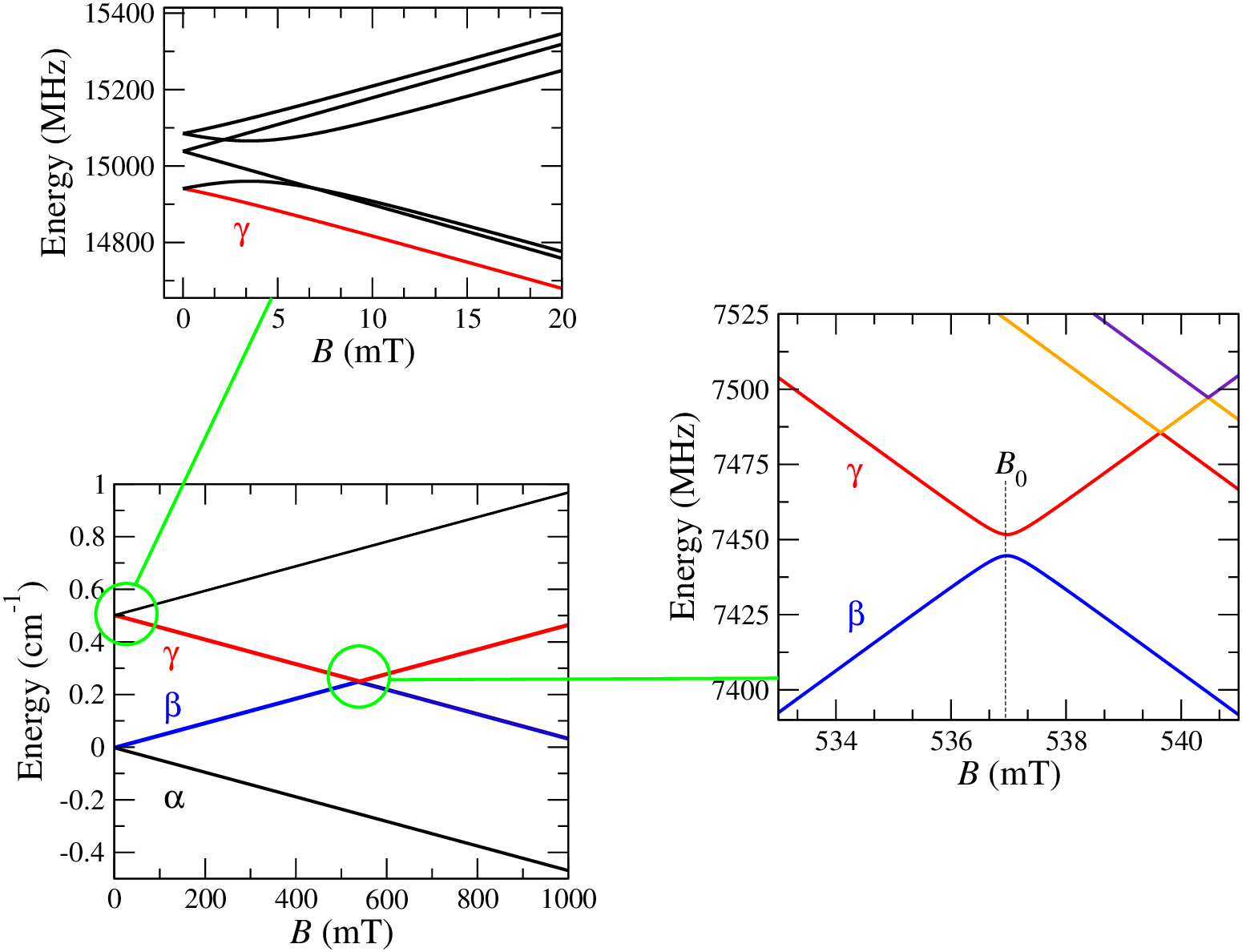}
\end{center}
\caption{Energy levels of the SrF($^2\Sigma$) molecule as functions of a magnetic field in the presence of an electric field of 1 kV/cm. The rotational constant of SrF is 0.251 cm$^{-1}$, the spin-rotation interaction constant $\gamma_{\rm SR}$ is $2.49\times10^{-3}$ cm$^{-1}$ and the dipole moment is 3.47 Debye. States $\beta$ and $\gamma$ undergo an avoided crossing at  the magnetic field value $B_{0}$. The value of $B_0$ varies with the electric field.}
\end{figure}

At zero electric field, the magnitude of the rotational angular momentum ${\bm N}$ is conserved. 
When ${\bm N}=0$, the spin-rotation interaction vanishes and two lowest energy levels of the molecule in a weak magnetic field correspond to the projections $M_S=-1/2$ and $M_S=+1/2$ of the spin angular momentum ${\bm S}$ on the magnetic field axis. The state with $M_S=-1/2$ (state $\alpha$ in Figure 1) is the absolute ground state of the molecule at all magnitudes of the magnetic field. The state with $M_S=+1/2$ (state $\beta$ in Figure 1) becomes degenerate with a high-field-seeking Zeeman state (state $\gamma$ in Figure 1) of the $N=1$ manifold at some value of a magnetic field. For SrF, this degeneracy occurs at the magnetic field $535.3$ mT.  States $\beta$ and $\gamma$ have different parity.  The ${\bm E}\cdot {\bm d}$ interaction is the only term in Hamiltonian (\ref{e2}) that couples states of different parity. Therefore, the crossing between states $\beta$ and $\gamma$ is real in the absence of an electric field and becomes avoided in the presence of an electric field. The properties of $^2\Sigma$ molecules, such as alignment and orientation, are very sensitive to electric and magnetic fields near these avoided crossings \cite{bretislav,timur-prl2006,erik-jcp2006}. In the present work, we show that the avoided crossings depicted in Figure 1 can be exploited for inducing and controlling collective spin excitations of $^2\Sigma$ molecules on an optical lattice.

\subsection{Magnetic excitons induced by electric fields}

We consider an optical lattice of SrF molecules prepared in the absolute (vibrational, rotational and Zeeman) ground state with one molecule per lattice site. We assume that the lattice sites are separated by 400 nm and the tunneling of molecules between lattice sites is suppressed, i.e. the molecules are in a Mott insulator state. This can be achieved by applying laser fields of high intensity \cite{Danzl2010}. With the current technology, it is possible to create optical lattices that provide harmonic confinement for ultracold atoms and molecules with the vibrational frequencies of the translational motion up to 100 - 150 kHz \cite{Bloch2005}. We assume that the molecules populate the ground state of the lattice potential and neglect the center-of-mass motion of the molecules  that broadens the spectral lines of the molecular crystal by about 5 $\%$ \cite{felipe}. 

The Hamiltonian describing the optical lattice with identical $^2\Sigma$ molecules in the presence of superimposed electric and magnetic fields is

\beq
\hat{H}=\sum_{n=1}^{N_{\text{mol}}} \hat{H}_{{\rm as}, n}
+\frac{1}{2}\sum_{n=1}^{N_{\text{mol}}}\sum_{m\neq n}^{N_{\text{mol}}}\hat{V}_{dd}({\bm r}_n-{\bm r}_m),
\label{lattice-hamiltonian}
\eeq
where ${\bm r}_n$ is the position of the {\it n}-th lattice site, $\hat{H}_{{\rm as}, n}$ is Hamiltonian (\ref{e2}) for molecule in site $n$,
 $\hat{V}_{dd}$ is the electric dipole-dipole interaction between molecules in different lattice sites, and $N_{\text{mol}}$ is the total number of molecules. 
Hamiltonian (\ref{lattice-hamiltonian}) can be rewritten in terms of the operators $\crea{B}{n}{f}$ and $\anni{B}{n}{f}$ that create and annihilate a molecular excitation $f$ in site $n$ as follows \cite{Agranovich:2008}: 

\begin{equation}
\label{H-full}
\hat{H}=\sum_{f} \hat{H}_f + \sum_{f} \sum_{g \neq f} \sum_{n}\sum_{m \ne n}J^{f,g}_{n,m}\crea{B}{n}{g}\anni{B}{m}{f},
\end{equation}

\noindent
where

\begin{equation}
\label{H-exciton}
\hat{H}_f = \epsilon _{f}\sum_{n} \crea{B}{n}{f}\anni{B}{n}{f} + \sum_{n}\sum_{m \ne n}J^{f,f}_{n,m}\crea{B}{n}{f}\anni{B}{m}{f},
\end{equation}

\noindent
$\epsilon_f$ is the energy of the $f$-th excited state of the molecule and 

\beq\label{J}
J^{f,g}_{n,m} =  \langle \phi_f ({\bm r_n})  \phi_0 ({\bm r_m}) | \hat{V}_{dd}({\bm r}_n-{\bm r}_m) | \phi_0 ({\bm r_n})  \phi_g ({\bm r_m}) \rangle
\eeq

The eigenstates of Hamiltonian (\ref{H-exciton}) can be written as 

\beq\label{exciton-states}
\Psi^f = \frac{1}{\sqrt{N_{\rm mol}}} \sum_{i} a_i \Phi^f_i
\eeq 

\noindent
where the wave functions

\beq\label{direct-products}
\Phi^f_i = \phi_f ({\bm r_i}) \prod_{j \neq i} \phi_{0} ({\bm r_j})
\eeq

\noindent
describe the ensemble of $N_{\rm mol} - 1$ molecules in the ground state $\phi_0$ and one molecule located at ${\bm r}_i$ in the $f$-th excited state. Eq. (\ref{exciton-states}) gives the wave function for a Frenkel exciton corresponding to an isolated $f$-th excitation of the molecule \cite{Agranovich:2008}. In general, different excitonic states are coupled by the terms $J^{f,g}_{n,m}$ with $f \neq g$ in Eq. (\ref{H-full}) and the eigenstate of Hamiltonian (\ref{lattice-hamiltonian}) is a linear combination of the excitonic states (\ref{exciton-states}) associated with different excitations $f$.

Hamiltonian (\ref{H-exciton}) can be diagonalized by the unitary transformation \cite{Agranovich:2008}

\beq
\crea{B}{n}{f} = \frac{1}{\sqrt{N_{\text{mol}}}}\sum_{\k}\crea{B}{\k}{f}e^{-i\k\cdot{\bm r}_n}
\eeq

\noindent
to yield 

\beq
\hat{H}_f= \sum_{\k}\left[\epsilon_f + L_f(\k)\right]\crea{B}{\k}{f}\anni{B}{\k}{f},
\label{exciton hamiltonian 2}
\eeq

\noindent
where $\k$ is the exciton wavevector and $L_{f}(\k)=\sum_n J^{f,f}_{n,0 }e^{i\k\cdot{\bm r}_n}$ determines the exciton dispersion in the absence of inter-exciton couplings. If all off-diagonal couplings $J^{f,g}_{n,m}$ with $f \neq g$ are much smaller that the energy separation of the molecular state $f$ from other molecular states, the coupling terms $J_{n,m}^{f,g}$ in Eq. (\ref{H-full}) can be neglected and Eq. (\ref{exciton-states}) provides an accurate description for the wave function of exciton $f$. This corresponds to the two-level approximation \cite{Agranovich:2008,Agranovich:2000}. 

In the present work, we consider the lowest energy excitation of the molecular crystal corresponding to the Zeeman transition from state $\alpha$ to state $\beta$ depicted in Figure 1. In the absence of electric fields and at low magnetic fields $B \ll B_0$, state $\alpha$ is $|N=0\rangle |M_S = -1/2 \rangle$ and state $\beta$ is $|N=0\rangle |M_S = +1/2 \rangle$. Because the electric dipole - dipole interaction is spin-independent, integral (\ref{J}) for $\phi_0 = |N=0, M_S = -1/2 \rangle$ and $\phi_f = |N=0, M_S=+1/2 \rangle$ vanishes. The $|N=0, M_S = -1/2 \rangle \rightarrow | N=0, M_S = + 1/2 \rangle$ excitation therefore produces a single isolated molecule in the excited state. Near the avoided crossing shown in Figure 1, state $\beta$ is a linear combination of $|N=0, M_S=-1/2 \rangle$ and  $|N=1, M_S=1/2 \rangle$ and the coupling constants $J^{f,f}_{n,m}$ must be non-zero. The matrix elements $J^{f,f}_{n,m}$ determine the exciton bandwidth defined as the difference $\Delta_f = |L_f(\pi/a) - L_f(0)|$, where $a$ is the intermolecular separation. Figure 2 shows the dominant nearest-neighbour coupling constant $J = J^{f,f}_{n, n+1}$ and the corresponding exciton bandwidth as functions of the magnetic field for three values of the electric field for a one-dimensional array of SrF molecules on an optical lattice. As the magnetic field is increased through the avoided crossing, state $\beta$ changes from predominantly $|N=0, M_S=1/2 \rangle$ to predominantly $|N=1, M_S=-1/2 \rangle$. As a result, the coupling constants $J_{n,m}^{f,f}$ giving rise to the lowest energy exciton as well as the exciton bandwidth increase from zero to a finite value. The variation of the coupling constants and the bandwidth with the magnetic field is slower for higher electric field.

The results presented in Figure 1 indicate that the first excited state of the SrF molecule is separated from other excited states by more than several MHz. Figure 2 demonstrates that 
the magnitudes of the coupling constants $J^{f,f}_{n,m}$ and the bandwidth of the lowest energy exciton are  $<$ 20 kHz. The off-diagonal matrix elements $J^{f,g}_{n,m}$ have similar magnitude as $J^{f,f}_{n,m}$. The results of Figures 1 and 2 thus indicate that the lowest energy exciton is well separated from the other molecular transitions so it should be well described by a single exciton wave function of the form (\ref{exciton-states}).  To prove this we calculated the energy band of the lowest energy exciton by the direct diagonalization of Hamiltonian (\ref{H-full}) with four coupled excitonic states and by the diagonalization of Hamiltonian (\ref{H-exciton}), which corresponds to the two-level approximation \cite{Agranovich:2008}.  Figure 2 demonstrates that the two-level approximation for the lowest energy exciton in the system considered here is very accurate. 

\begin{figure}[ht]
\label{J-and-Delta}
\begin{center}
\includegraphics[scale=0.40]{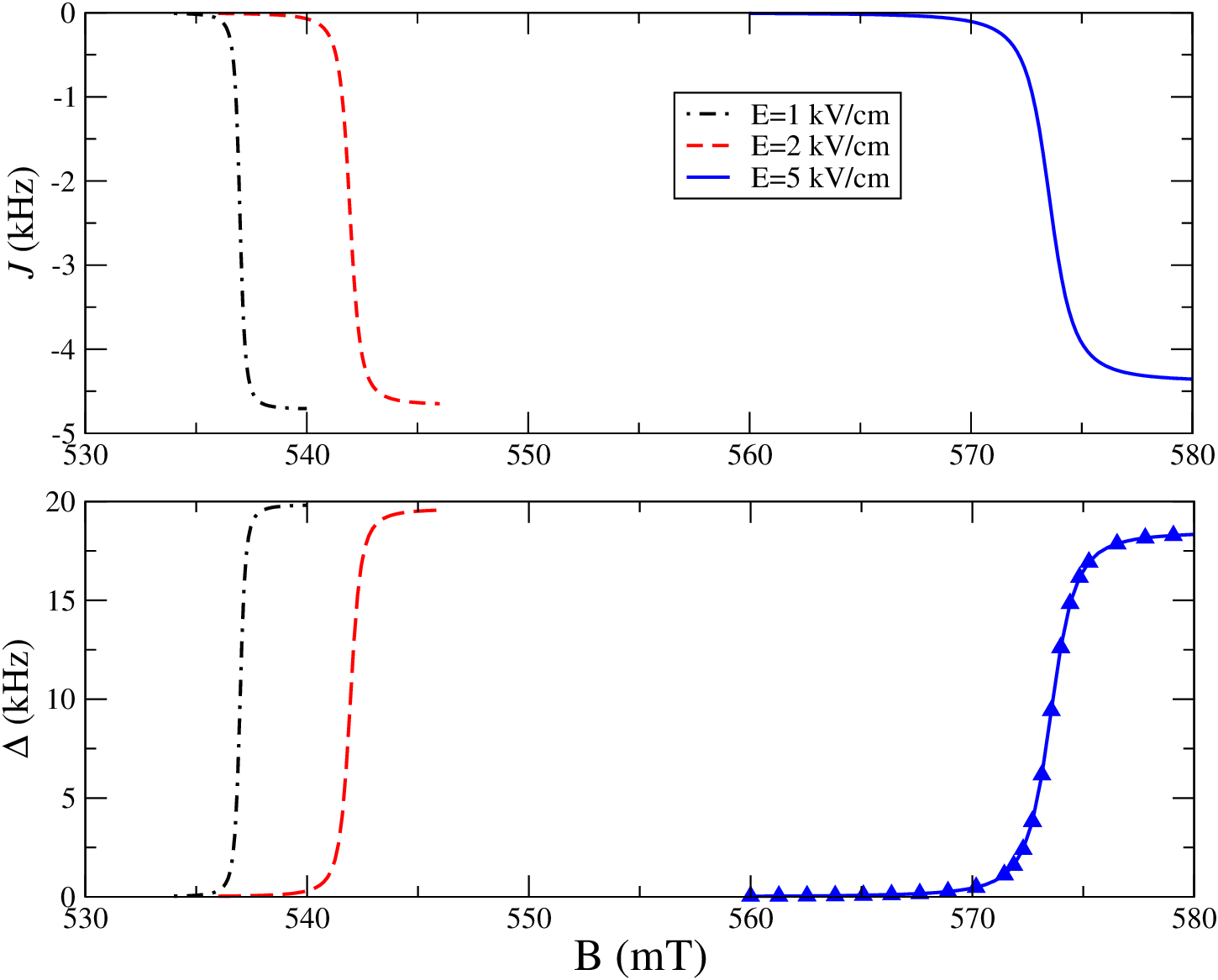}
\end{center}
\caption{Upper panel: Nearest-neighbour coupling constants $J = J^{f,f}_{n,n+1}$ (upper panel) giving rise to the lowest energy exciton in a crystal of $^2\Sigma$ molecules calculated for three values of the electric field as functions of the magnetic field. Lower panel: Exciton bandwidths $\Delta$ calculated for three values of the electric field as functions of the magnetic field. The calculations are for 2000 SrF($^2\Sigma$) molecules arranged in a one-dimensional array with the lattice spacing 400 nm.  Lines in the lower panel represent the results obtained by the diagonalization of Hamiltonian (\ref{H-full}) with four coupled excitons. Symbols show the results of the calculation using the two-level approximation for the lowest energy excitation. 
}
\end{figure}

Figure 2 suggests an interesting possibility of creating an entangled many-body state (\ref{exciton-states}) of {\it non-interacting} molecular spins. 
This can be achieved by the following procedure:  (i) apply an electric field to couple the molecular states of different parity; (ii) tune the magnetic field adiabatically to a value $B > B_0$ (see Figure 1), where the $\alpha \rightarrow \beta$ exciton bandwidth is large; (iii) generate the lowest energy excitation; (iv) detune the magnetic  field to a value $B \ll B_0$ and turn off the electric field. As mentioned in the previous section, when the electric field is absent and the magnetic field is $B \ll B_0$, state $\alpha$ is a pure $M_S=-1/2$ state and state $\beta$ is a pure $M_S=+1/2$ state. If carried out faster than the inherent time scale of the exciton dynamics (see next section),  step (iv) must therefore project the excitonic wave function on the many-body state

\beq\label{spin-exciton}
\Psi = \frac{1}{\sqrt{N_{\rm mol}}} \sum_{i} C_i \Phi^S_i
\eeq 

\noindent
with 

\beq\label{spin-functions}
\Phi^S_i = |M_S=1/2\rangle_{{\bm r}_i} \prod_{j \neq i} |M_S=-1/2\rangle_{{\bm r}_j}. 
\eeq

\noindent
The variation of the magnetic field in step (iv) must be generally faster than $h/J$ to preserve the exciton state, but slow enough to preclude the non-adiabatic transitions to molecular state $\gamma$. This can be achieved because the splitting of the molecular states $\beta$ and $\gamma$ at the avoided crossing is much greater than $J$. We have confirmed that the magnetic field can be detuned to a value $B \ll B_0$ without changing the magnitudes of the coefficients $C_i$ by time-dependent calculations as described in the next section. Wave function (\ref{spin-exciton}) is a many-body analogue of the two-body EPR state with spin-$1/2$ particles $a | \uparrow \rangle | \downarrow \rangle + b | \downarrow \rangle | \uparrow \rangle$ \cite{many-body-entanglement, book-on-quantum-information}. 
In the following section, we show that the expansion coefficients $C_i$ in Eq. (\ref{spin-exciton}) can be modified by creating vacancies in the molecular crystal.

\subsection{Localization of magnetic excitons in the presence of vacancies}

In an ideal, infinitely large crystal, the exciton wave function is completely delocalized and the probability to find any molecule in the excited state is the same. If the molecular crystal contains impurities in the form of other molecules or vacancies, the exciton wave function is modified and the excitons may undergo coherent localization \cite{Agranovich:2008}. The localization patterns depend on the dimensionality of the crystal, the concentration and the distribution of impurities as well as the exciton - impurity interaction strength. 
Wurtz and coworkers have recently presented the results of an experiment showing that atoms in specific sites of an optical lattice can be selectively evaporated by focusing an electron beam onto a particular lattice site \cite{vacancies}. This technique is general and can be applied 
 to molecules as well as atoms. This method can be used to create mesoscopic ensembles of ultracold particles with arbitrary spatial patterns. 
In the following two sections, we show that the interactions of magnetic excitons (\ref{spin-exciton}) with empty lattice sites are very strong and that the entangled spin states given by Eq. (\ref{spin-exciton}) can be effectively modified by evaporating molecules from the optical lattice to create vacancies. In this section, we consider a one-dimensional array of 2000 SrF molecules on an optical lattice with lattice spacing $a=400$ nm. Since the exciton coupling constant $J$ is negative ({\it cf}, Figure 2),  the exciton has a positive effective mass, i.e. the energy of the exciton increases with the wavevector \cite{Agranovich:2008}. An excitation of molecules by laser field produces excitons with small wavevector \cite{Agranovich:2008} so we focus on a state near the bottom of the energy spectrum.

\begin{figure}[ht]
\label{localization}
\begin{center}
\includegraphics[scale=0.55]{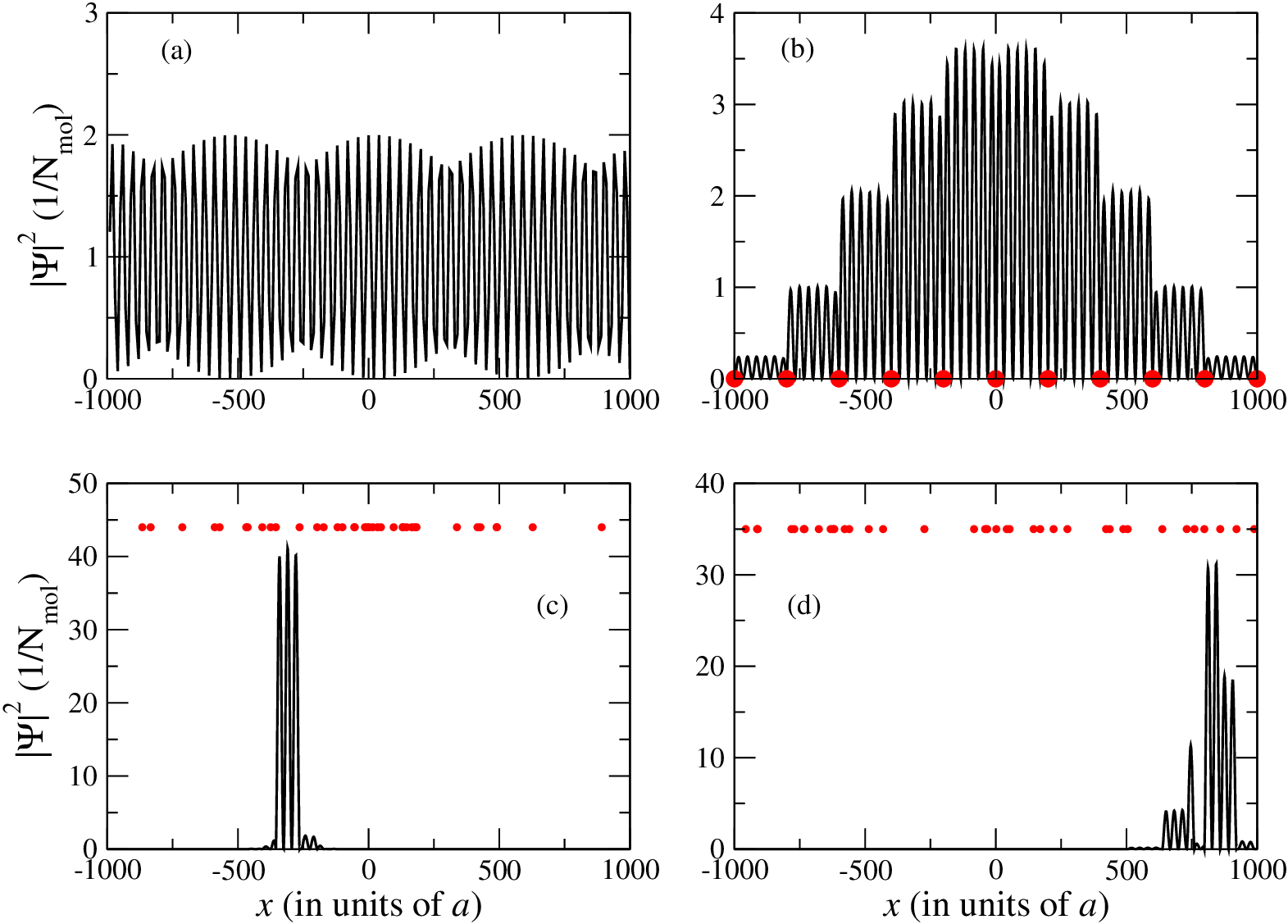}
\end{center}
\caption{Magnetic exciton wave function (\ref{spin-functions}) for a one-dimensional array of SrF($^2\Sigma$) molecules in an optical lattice with 2000 lattice sites separated by $a=400$ nm. Panel (a): Free exciton wave function. Panel (b): Exciton wave function in the presence of equally spaced vacancies. Panel (c): Exciton wave function in the presence of a Gaussian distribution of randomly placed vacancies. Panel (d): Exciton wave function in the presence of a uniform distribution of randomly placed vacancies. The number of vacancies for Panels (c) and (d) is 37. The positions of the vacancies in Panels (b), (c) and (d) are indicated by the symbols.
The wave function is presented for state 51 in the spectrum of $\sim$ 2000 states. The magnetic field is 536.9 mT and the electric field is 1 kV/cm.}
\end{figure}

To find the eigenstates of a molecular crystal with vacancies, we use the two-level approximation and modify Hamiltonian (\ref{H-exciton}) and the basis wave functions (\ref{spin-functions}) to omit the terms corresponding to the positions of the empty lattice sites. The two-level approximation was shown to be very accurate in the previous section. The dimension of the Hamiltonian matrix is ($N_{\rm mol} - N_{\rm vac}$)$\times$($N_{\rm mol} - N_{\rm vac}$), where $N_{\rm vac}$ is the number of vacancies in the lattice. We calculate the eigenvectors and eigenvalues using two different numerical methods in order to verify the accuracy of the computations. Figure 3 shows the probability density of the magnetic exciton for the array of SrF molecules with no defects (panel a), with vacancies positioned at every $200$-th site (panel b), with a Gaussian distribution of 37 randomly placed vacancies (panel c) and with a uniform distribution of 37 randomly placed vacancies (panel d). The results show that the presence of vacancies modifies the exciton wave function to a great extent and leads to various localization patterns of the exciton wave functions. In particular, a periodic placement of vacancies (Figure 2b) leads to the formation of wave function domains with the molecules in the middle of the crystal being much more likely in the $M_S=+1/2$ state. The Gaussian and uniform distributions of impurities lead to the formation of a strongly localized state with only a few molecules sharing the $M_S=+1/2$ excitation. The localization patterns can be further modified by varying the concentration of impurities, which is demonstrated in Figure 4.

\begin{figure}[ht]
\label{localization-percentage}
\begin{center}
\includegraphics[scale=0.55]{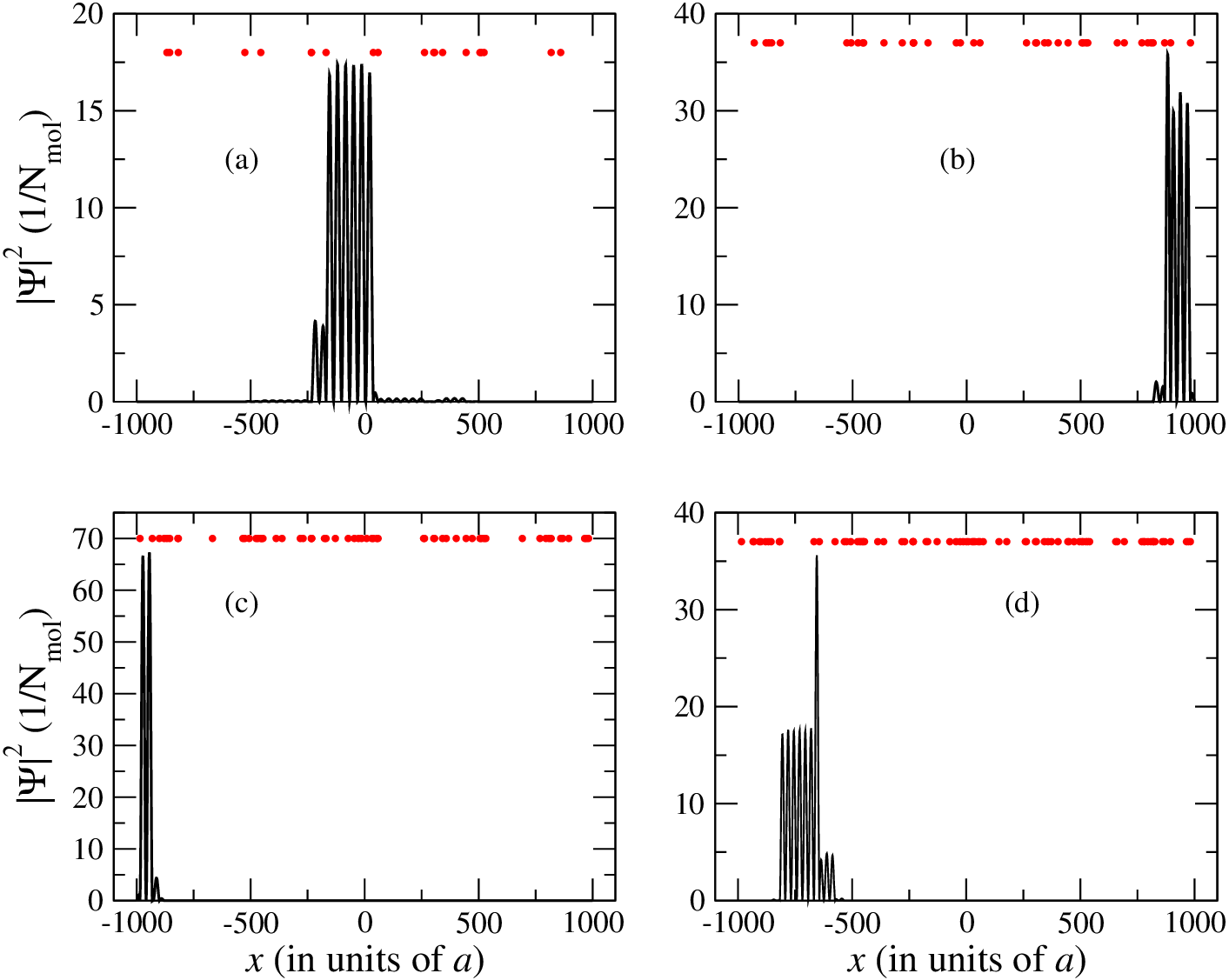}
\end{center}
\caption{Magnetic exciton wave function (\ref{spin-functions}) for a one-dimensional array of SrF($^2\Sigma$) molecules in an optical lattice with 2000 lattice sites separated by $a=400$ nm in the presence of a uniform distribution of randomly placed vacancies. Different panels correspond to different concentrations of vacancies. Panel (a): 20 vacancies; panel (b): 40 vacancies; panel (c): 60 vacancies; panel (d): 80 vacancies. The positions of the vacancies are indicated by the symbols. The magnetic field is 536.9 mT and the electric field is 1 kV/cm.}
\end{figure}

\subsection{Dynamics of spin excitation transfer}

The system proposed in the present work may allow for the study of the dynamics of spin excitation transfer between molecules in real time. A gradient of a magnetic field applied to an optical lattice of molecules initially in the $M_S=-1/2$ state can be used to excite a molecule in a specific lattice site to the $M_S=+1/2$ state by an rf $\pi/2$ pulse resonant with the $M_S=-1/2 \rightarrow M_S=+1/2$ transition for the molecule in this particular position \cite{demille}. This should produce a spin excitation localized on a single molecule. The spin excitation transfer can then be induced by removing the field gradient, tuning the magnetic field to a value $B \sim B_0$ and applying an electric field that gives rise to significant couplings $J^{f,f}_{n,m}$ in Eq. (\ref{H-exciton}). The spin excitation of a single molecule is a localized exciton wave packet. When the coupling matrix elements $J^{f,f}_{n,m}$ are non-zero, the wavepacket must exhibit dynamics of spin excitation transfer  determined by the strength of the coupling matrix elements $J^{f,f}_{n,m}$. 


In order to explore the effects of electric and magnetic fields on the dynamics of the exciton wavepackets and determine the timescale of the spin excitation transfer, we consider a small ensemble of SrF molecules on a one-dimensional optical lattice. 
In a finite-size ensemble of molecules, the spin excitation should propagate to the edge of the molecular sample and return to the original molecule, leading to revivals of the exciton wavepacket.
We expand the total wave function of the system in terms of direct products (\ref{spin-functions}) with time-dependent expansion coefficients, 

\begin{equation}
\label{time-decomp}
\Psi =  \sum_{n} F_{n}(t) \Phi^S_{n}.
\end{equation}

\noindent
The substitution of this expansion in the time-dependent Schr\"{o}dinger equation with Hamiltonian (\ref{H-exciton}) yields a system of first-order coupled differential equations for the expansion coefficients $F_n(t)$. We integrate these equations to obtain the time evolution of the quantities $|F_n(t)|^2$, which give the probability for the spin excitation to be found at time $t$ in lattice site $n$. Figure 5 shows the 
autocorrelation function $|A(t)|^2 = |\langle \Psi(t=0)| \Psi(t) \rangle|^2$, which describes the 
probability for the spin excitation to be on the initially excited molecule (molecule four in this example) in a system of seven molecules calculated as a function of time for four different magnitudes of the magnetic field near the avoided crossing shown in Figure 1. The electric and magnetic fields are kept constant in this calculation. Figures 5 and 6 demonstrate that the spin excitation is transferred to other molecules and then comes back to the original site at regular time intervals. The frequency of the revivals is very sensitive to the magnetic and electric fields and varies from $\sim 1$ to $\sim 5$ ms.

Vacancies introduce potential barriers which influence the dynamics of excitons. The strength of the barriers can be modified by producing several vacancies in adjacent lattice sites. Figure 7 demonstrates the interaction of the spin excitation with the potential barriers produced by two (panel a) and three (panel b) adjacent vacancies. This calculation shows that the excitation can tunnel through the vacancies and that the rate for tunneling through three empty lattice sites is almost one order of magnitude smaller than that for tunneling through two empty sites. Figure 7 demonstrates that the spin excitation in a mesoscopic ensemble of $^2\Sigma$ molecules in an optical lattice separated by three empty lattice sites can be considered isolated on the time scale of $<$ 5 ms.

\begin{figure}[ht]
\label{spin-excitaton-transfer-B}
\begin{center}
\includegraphics[scale=0.55]{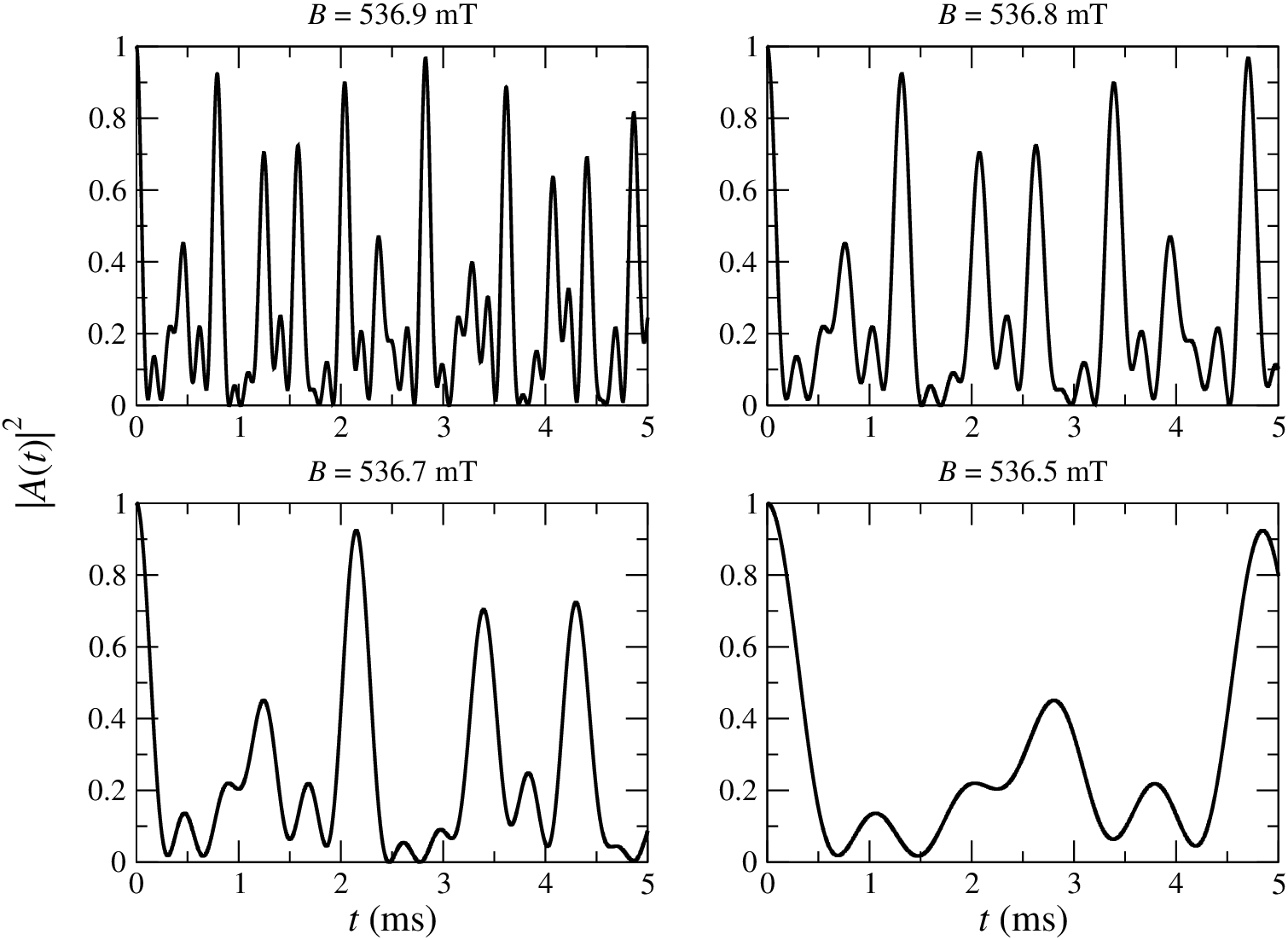}
\end{center}
\caption{Time-dependent probability of the spin excitation to be localized on molecule four in a one-dimensional array of seven molecules in an optical lattice with lattice spacing $a = 400$ nm for different magnetic fields near the avoided crossing shown in Figure 1. The electric field magnitude is 1 kV/cm.
}
\end{figure}

\begin{figure}[ht]
\label{spin-excitaton-transfer-E}
\begin{center}
\includegraphics[scale=0.55]{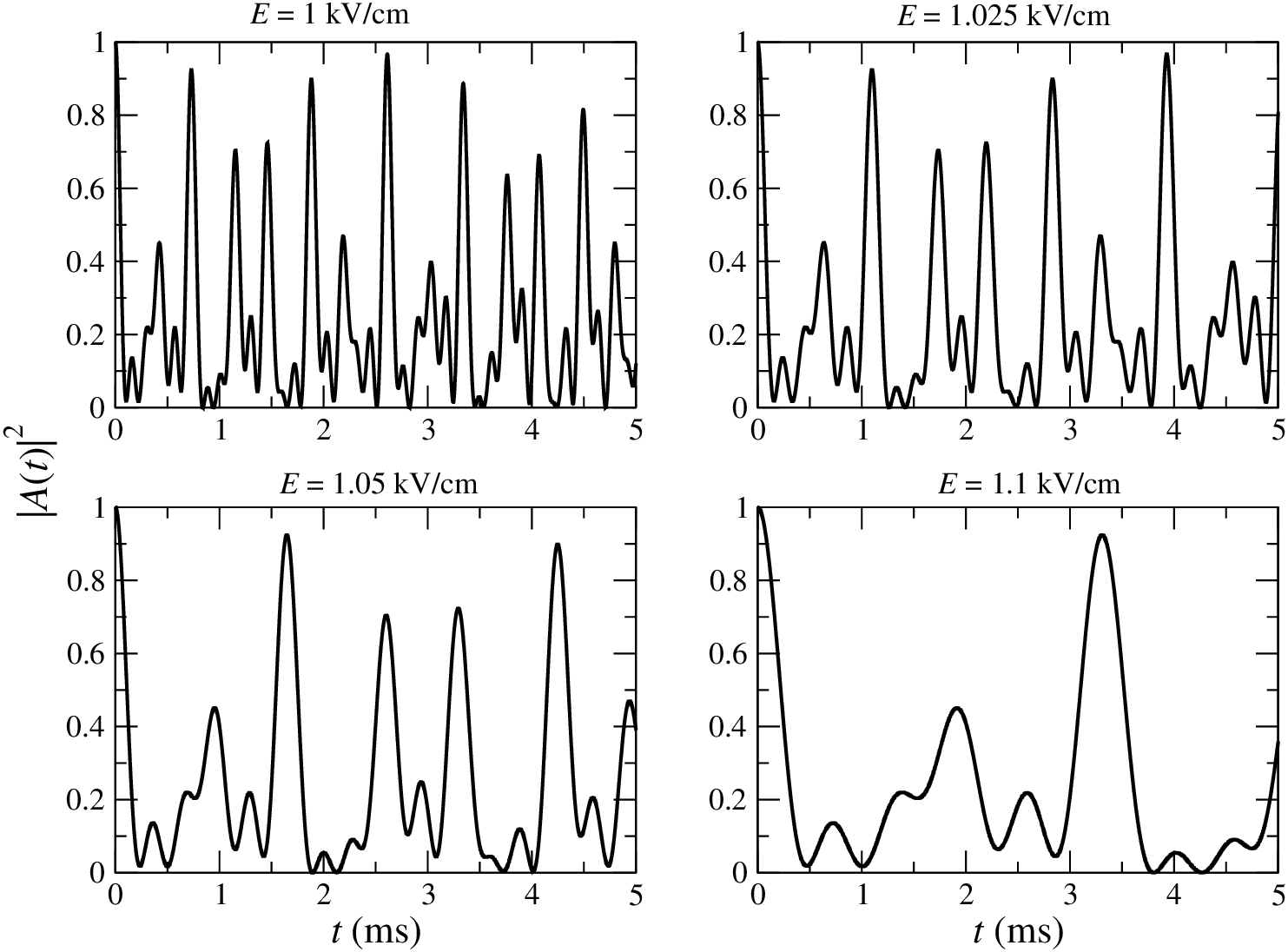}
\end{center}
\caption{Time-dependent probability of the spin excitation to be localized on molecule four in a one-dimensional array of seven molecules in an optical lattice with lattice spacing $a = 400$ nm for different electric fields near the avoided crossing shown in Figure 1. The magnetic field magnitude is 536.9 mT.
}
\end{figure}

\begin{figure}[ht]
\label{tunneling}
\begin{center}
\includegraphics[scale=0.55]{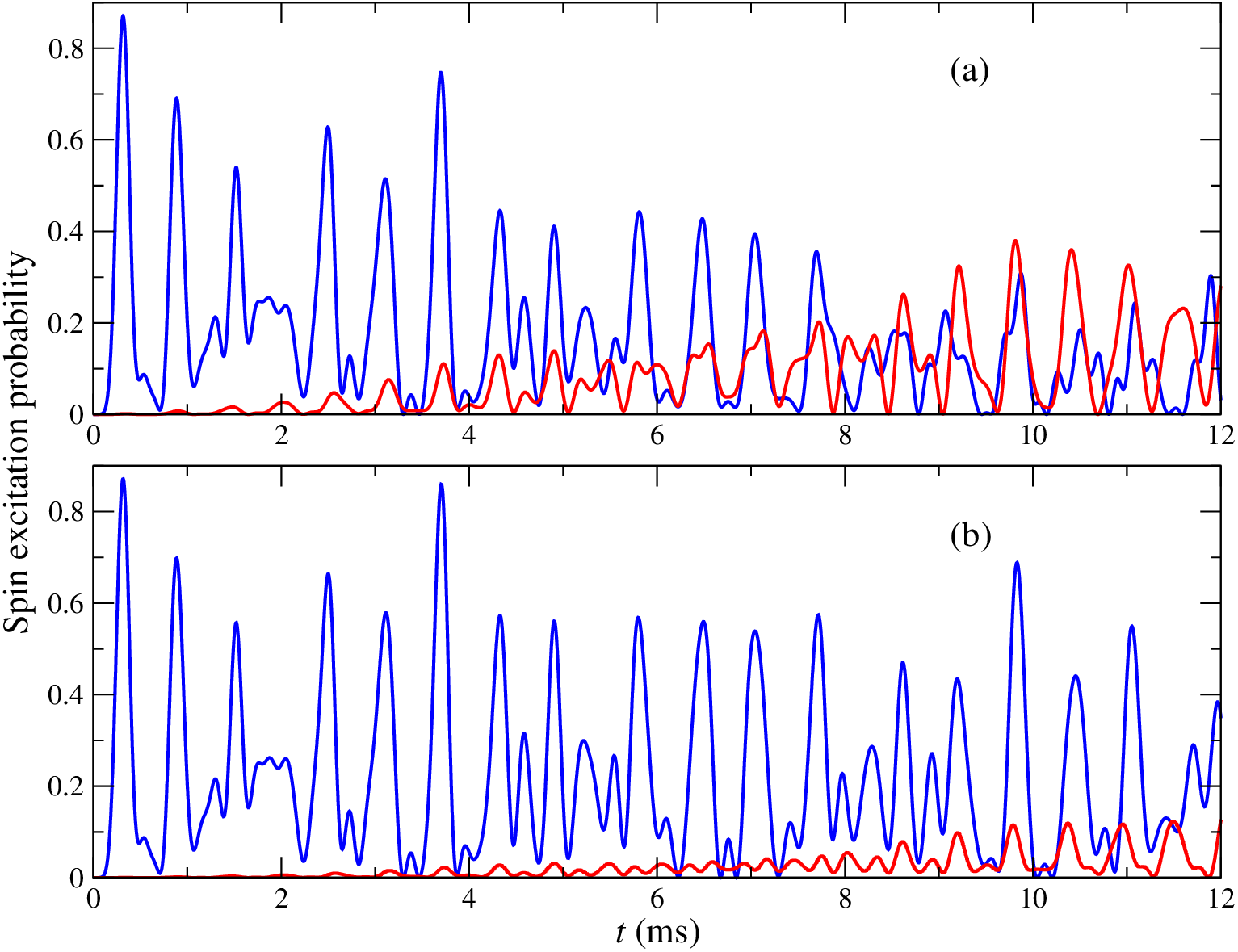}
\end{center}
\caption{Time-dependent probability of the spin excitation to be localized on molecule five (solid blue line) and molecule 6 (solid red line) in a one-dimensional array of ten molecules in an optical lattice with lattice spacing $a = 400$ nm. Panel (a): molecules five and six are separated by two empty lattice sites. Panel (b): molecules five and six are separated by three empty sites.  The magnetic field magnitude is 536.9 mT and the electric field is 1 kV/cm. At $t=0$, the spin excitation is localized on molecule four.
}
\end{figure}

\section{Summary}

We have shown that the unique energy-level structure of $^2 \Sigma$ molecules can be exploited for controlled preparation of many-body entangled states (\ref{spin-exciton}) of non-interacting molecular spins on an optical lattice.
The proposed procedure is based on inducing the interactions giving rise to the formation of magnetic excitons and projecting the exciton states on the basis of uncoupled spin states by varying external electric and magnetic fields. Our calculations show that the exciton states are formed at 
particular combinations of electric ($E_0$) and magnetic ($B_0$) fields  near the avoided crossing between states $\beta$ and $\gamma$ depicted in Figure 1. The molecular spin states become uncoupled at magnetic fields $B \ll B_0$. To confirm that the detuning of the magnetic field does not affect the exciton states, we performed a time-dependent calculation with seven molecules prepared at $t=0$ in the eigenstate of Hamiltonian (\ref{H-exciton}) at the magnetic and electric fields near $B_0$ and $E_0$. The calculation showed that the exciton wave function remained the same upon detuning of the magnetic field to a value $B \ll B_0$. This result is largely independent of the speed of the magnetic field variation because the exciton state, once formed, remains the eigenstate of  Hamiltonian (\ref{H-exciton}) at all values of the magnetic field. It may also be interesting to generate many-body entangled states by first creating a spin excitation on an isolated molecule and then tuning the magnetic and electric fields close to $B_0$ and $E_0$. This should produce a dynamically changing exciton wavepacket. 
In order to elucidate the feasibility of projecting the exciton wavepackets on the basis of uncoupled spin states by detuning the magnetic and electric fields from $B_0$ and $E_0$, we calculated the time-evolution of the magnetic exciton wavepackets at electric and magnetic fields near the values $E_0$ and $B_0$.  Our calculations indicate that the time scale for the magnetic excitation transfer at magnetic and electric fields near $E_0$ and $B_0$ is  $ \sim 1$ ms.  Our results show that it is sufficient to change the electric field by 0.1 kV/cm and the magnetic field by 5 Gauss to suppress significantly the time evolution of the exciton wavepackets. This variation of the dc magnetic and electric fields can be achieved on the time scale of $ \sim 1$ $\mu$s \cite{Bethlem2000,Chin2010}. 

We use SrF molecules for our illustrative calculations in this paper. SrF has a relatively large dipole moment, a relatively small rotational constant and a very weak spin-rotation interaction by comparison with other $^2\Sigma$ molecules \cite{Egorov2004}. The effects studied here should be more pronounced, making the experiments easier, in an ensemble of molecules with a larger dipole moment and a larger spin-rotation interaction constant or with an optical lattice with smaller lattice-site separations. The experimental work on the creation of ultracold $^2\Sigma$ molecules that should exhibit similar behavior as SrF is currently underway in several laboratories \cite{doyle-caf,Shuman2009,dirosa,Hudson2002}. We note that molecules in the $^3\Sigma$ electronic state exhibit similar avoided crossings in combined electric and magnetic fields \cite{timur2006,hutson}, which significantly widens the range of molecules that can be used for the experiments proposed here.

 The absence of interactions between the spin states in the coherent superpositions (\ref{spin-exciton}) should make the entanglement robust against decoherence due to vibrational motion of molecules in the lattice. We note that this entanglement is created over distances spanning thousands of molecules in an optical lattice separated by 400 nm, which amounts to mm size scales.
  We showed that the exciton states can be localized by creating vacancies in the optical lattice. Our results demonstrate that the localization patterns of the magnetic exciton states are sensitive to the number and distribution of vacancies. This can be exploited for engineering entangled states with different patterns.   
   Excitons determine the optical properties of crystals and it might also be interesting to explore the interaction of the magnetic excitations with electromagnetic radiation, which can be achieved either through coupling to a two-photon optical field \cite{Lukin2001} or to a single photon rf field \cite{Tscherbul2010}. The localization of excitons by vacancies in the presence of tunable dc electric and magnetic fields can then be used for controlled preparation of polaritons. The experimental techniques for producing optical lattices of ultracold molecules in the ro-vibrationally ground states \cite{Danzl2010} and for creating vacancies in optical lattices  with controlled spatial arrangements \cite{vacancies} have been recently demonstrated.

The system proposed here can be used to study dynamics of spin excitation transfer in molecular aggregates. We have shown that localized magnetic excitations in finite-size arrays of molecules exhibit revivals in the presence of electric and magnetic fields near $E_0$ and $B_0$. The frequency of the revivals can be controlled by varying the magnitude of the electric and magnetic fields. Our calculations show that the magnetic excitation can tunnel through multiple vacancies in adjacent lattice sites. The tunneling rate is dramatically suppressed when the number of vacancies in adjacent lattice sites is increased.  Our results indicate that the time scale for the magnetic excitation transfer through three adjacent vacancies is $> 5$ ms. This suggests a possibility of studying spin excitation transfer by measuring the exciton wavepacket revivals in small molecular aggregates. The localized spin excitations in an optical lattice of $^2\Sigma$ molecules can be created and observed by applying a gradient of a magnetic field and performing a spatially-resolved spectroscopic measurement \cite{demille,Bakr2009}. 

\section{Acknowledgments}
We acknowledge useful discussions with Dr. Marina Litinskaya. The work was supported by NSERC of Canada and the Peter Wall Institute for Advanced Studies at the University of British Columbia. JPR acknowledges the support of Prof. M. I. Hern\'{a}ndez and Prof. J. Campos-Mart\'{i}nez and the Krems group for hospitality during the visit to the University of British Columbia. The work of JPR was financially supported by 
a JAE-pre fellowship from CSIC.

\end{document}